\documentstyle[12pt]{article}
\begin{document}
\title {Up--dating for the paper ``The mass of the Higgs boson in the
Standard Model from precision tests'' (MPI--Ph/95--39, hep--ph/9505304).
\thanks{Supported in part by the Polish Committee for Scientific
        Research and European Union Under contract CHRX-CT92-0004.}}

\author{Piotr H. Chankowski \\
Institute of Theoretical Physics, Warsaw University\\
ul. Ho\.za 69, 00--681 Warsaw, Poland.\\
\\
Stefan Pokorski
\thanks{On leave of absence from
Institute of Theoretical Physics, Warsaw University}
\\
Max--Planck--Institute f\"ur Physik\\
Werner -- Heisenberg -- Institute \\
F\"ohringer Ring 6,
80805 Munich, Germany
}

\maketitle


\begin{abstract}
The bounds on the Higgs mass in the Standard Model
are re-analyzed using the precision electroweak data presented
at the International Europhysics Conference on High Energy Physics,
Brussels, 27 July -- 2 August, 1995
\end{abstract}

\newpage

After submission of the paper \cite{MY} for publication, new data from the SLD
and LEP became available. We present here an up--date of our fits with
new measurements included. The experimental input used in the present
fits is summarized in ref \cite{BRUSS}.
The bulk of the LEP results show satisfactory agreement with the earlier
reports (one should note, however, the new value ~
$A_{FB}^{b,0}=0.0997\pm0.0031$). ~For ~
$M_W$ ~we use (as previously) ~$80.33\pm0.17$ GeV which is the average
value of the UA2 measurement and the new measurement reported by the
CDF \cite{MORIOND95} (the D0 collaboration has not published the results
of their new analysis yet). When the top quark mass is included in the fit
we use the value ~$m_t=181\pm12$ GeV. ~For ~$\Delta\alpha_{EM}^{hadr}$ ~we
use the value
\footnote{The result reported in the ref. \cite{SWA} has
been recently updated \cite{PES,BRUSS} and is now close to the results
reported in \cite{JEG}.
The result of ref. \cite{MAR} is about ~$1\sigma$ ~lower than that of
\cite{JEG}; it is, however, based on more theoretical assumptions
\cite{BRUSS}.}
\cite{JEG} ~$0.0280\pm0.0007$ ~
and include it into the ~$\chi^2$ ~fit.

The most significant changes in the LEP data are ~$R_b=0.2219\pm0.0017$, ~and ~
$R_c=0.1540\pm0.0074$. ~Since the identification of ~$b$ ~quarks is much better
than of the ~$c$ ~quarks experimental collaborations also
quote the value ~$R_b=0.2206\pm0.0016$ ~which is obtained when the
value of ~$R_c$ ~is fixed to its SM prediction ~$R_c=0.172$. ~

The new SLD result read \cite{SLD,BRUSS}: ~
$A_{LR}^e\equiv{\cal A}_e=0.1551\pm0.0040$ ~(which
corresponds to ~$\sin^2\theta^{eff}_l=0.23049\pm0.00050$). ~The SLD
collaboration also reported for the first time \cite{BRUSS} the results ~
$A_{LR}^b\equiv{\cal A}_b=0.841\pm0.053$ ~and ~
$A_{LR}^c\equiv{\cal A}_c=0.606\pm0.090$. ~

The new SLD data for ~$A_{LR}^e$~ and the LEP value still remain more than ~
$2\sigma$ ~apart: although the central values are now much closer to each
other, the SLD error has significantly decreased.

The newly reported results for ~$R_b$, ~$R_c$, ~$A_{LR}^b$, ~$A_{LR}^c$ ~
have drastic effects on SM fits whose quality has significantly decreased.
This indicates either large fluctuations in the present data or new
physics (or both). Here we adopt the first point of view and assume that
the SM is the correct low energy effective theory at the electroweak
scale. In this framework we critically re--examine the bounds on the Higgs
boson mass in several different versions of the fit.

We begin with the results of a fit to all the  LEP, SLAC and Fermilab data
described above (i.e. to ~$M_W$, ~$1-M^2_W/M^2_Z$, ~$\Gamma_Z$, ~
$\sigma_{h}$, ~${\cal A}_e$,  ~${\cal A}_{\tau}$, ~
$\sin^2\theta^{lept}_{eff}<Q_{FB}>$, ~${\cal R}_l$, ~$A_{FB}^{0,l}$, ~
$R_b$, ~$R_c$, ~$A_{FB}^{0,b}$, ~$A_{FB}^{0,c}$, ~$A_{LR}^e$, ~
$A_{LR}^b$, ~$A_{LR}^c$, ~$m_t$ ~and ~$\Delta\alpha^{hadr}_{EM}$). ~
They are given in the first line of Table 1 ~and ~$1\sigma$ ~and ~$2\sigma$ ~
bounds in the ~$(m_t, M_h)$ ~plane are shown in Fig.1.
Significant upper bounds on the Higgs boson mass are obtained but the quality
of the fit is  poor.

\begin{center}
{\bf Table 1.} Results of the fit to all the data
\cite{BRUSS,MORIOND95,SLD}. All masses in GeV.
\vskip 0.2cm
\begin{tabular}{||l|l|l|l|l|l|l|l||} \hline
$m_t$                 &
$\Delta m_t$          &
$M_h$                 &
$\Delta M_h$          &
$\alpha_s(M_Z)$       &
$\Delta\alpha_s(M_Z)$ &
$\chi^2$              &
d.o.f                 \\ \hline
171.0&11.1& 93&$^{+189}_{-63}$ &0.122&0.005&25.0&14\\ \hline
170.5&10.9& 82&$^{+181}_{-55}$ &0.122&0.005&21.1&12\\ \hline
\end{tabular}
\end{center}

We now examine the impact of the observables which give the dominant
contribution to ~$\chi^2$ ~on the Higgs boson mass bounds.
In the second line of Table 1 ~there are also given the results
of a fit without the two new SLAC measurements of ~$A_{LR}^b$ ~and ~
$A_{LR}^c$. ~The bounds on ~$M_h$ ~are slightly stronger and the fit
is a bit better.

In the first row of Table 2 ~
we show the results of a fit without ~$R_b$, ~$R_c$~ and without
all the ~$LR$ ~asymmetries measured by the SLD ~($A_{LR}^e$ ~and ~
$A_{LR}^b$, ~$A_{LR}^c$). ~The ~$\chi^2$ ~value is now excellent.
The ~$1\sigma$ ~and ~$2\sigma$ ~contours in the ~$(m_t, M_h)$ ~plane
are shown in Fig.2a. The bounds on ~$M_h$ ~are very weak and critically
depend on the value of  ~$m_t$ ~measured in Fermilab.
Without ~$m_t$ ~in the fit the included  data only correlate ~
$M_h$ ~with ~$m_t$ ~(as shown also in Fig.2a) with, however, no ~
$2[s]$ ~upper bound on ~$M_h$. ~

Another observable which depends on ~$m_t$ ~but not on ~$M_h$ ~is ~$R_b$. ~
The measured value, to be consistent with the SM, requires very low value
of ~$m_t$. ~Thus, inclusion of ~$m_t$ ~and/or ~$R_b$ ~gives the fitted value
of ~$m_t$ ~lower than the Fermilab value and, in consequence, stronger
bounds on ~$M_h$. ~This is also shown in Table 2 ~and in Fig. 2b.  ~Here we
give the results of fits with ~$R_b=0.2206\pm0.0016$ ~which is the
experimental number obtained under the assumption of ~$R_c=0.172$ ~
(with the new values of ~$R_b$ ~and ~$R_c$ ~we get very similar bounds
but the quality of the fit is poorer).

\begin{center}
{\bf Table 2.} Results of the fit without the SLD asymmetries.
\vskip 0.2cm
\begin{tabular}{||l|l||l|l|l|l|l|l|l|l||} \hline
$m_t$                 &
$R_b$                 &
$m_t$                 &
$\Delta m_t$          &
$M_h$                 &
$\Delta M_h$          &
$\alpha_s(M_Z)$       &
$\Delta\alpha_s(M_Z)$ &
$\chi^2$              &
d.o.f                 \\ \hline
$+$&$-$&180.1&11.9&335&$^{+529}_{-222}$&0.125&0.005&3.2&9\\ \hline
$-$&$+$&152.2&$^{+14.7}_{-12.9}$&52&$^{+115}_{-29}$&0.123&0.005&9.4&9\\ \hline
$+$&$+$&172.4&11.2&189&$^{+326}_{-122}$&0.124&0.005&11.6&10\\ \hline
\end{tabular}
\end{center}

As the next step we include in the fits the SLD value of ~$A_{LR}^e$. ~
Its effect on the fits can be understood from Fig. 2a. We show there
the contour of constant ~$A_{LR}^e=0.1551$. ~It is almost parallel to the
open contours which show the ~$m_t - M_H$ ~correlation without ~
$m_t$, ~$R_b$ ~and ~$A_{LR}^e $~in the fit
but for the same ~$m_t$ ~the correlated ~$M_h$ ~is much lower. It is clear
therefore that to a very good approximation the inclusion of ~$A_{LR}^e$ ~
should not alter the fitted value of ~$m_t$ ~but strenghten the bound on ~
$M_h$. ~This is indeed seen in Table 3 ~and in Fig. 3 ~(small changes in ~
$m_t$ ~compared to Table 2 ~are due to the fact that the
two discussed above contours in Fig.2 are not exactly parallel).

\begin{center}
{\bf Table 3.} Results of the fit as in Table 2 but with ~$A_{LR}^e$
included.
\vskip 0.2cm
\begin{tabular}{||l|l||l|l|l|l|l|l|l|l||} \hline
$m_t$                 &
$R_b$                 &
$m_t$                 &
$\Delta m_t$          &
$M_h$                 &
$\Delta M_h$          &
$\alpha_s(M_Z)$       &
$\Delta\alpha_s(M_Z)$ &
$\chi^2$              &
d.o.f                 \\ \hline
$+$&$-$&176.7&12.0&125&$^{+271}_{-89}$&0.122&0.005&8.9&10\\ \hline
$-$&$+$&155.6&11.1&33&$^{+49}$&0.122&0.005&14.8&10\\ \hline
$+$&$+$&169.6&10.7&79&$^{+160}_{-54}$&0.122&0.005&17.1&11\\ \hline
\end{tabular}
\end{center}

The third line in Table 3 ~ can be compared with the second line in Table 1 ~
to see the effect of using new values of ~$R_b$ ~and ~$R_c$ ~instead of the
value of ~$R_b=0.2206\pm0.0016$ ~obtained with ~$R_c$ ~fixed to its
SM value. We conclude that the limits on ~$M_h$ ~are stable with
respect to the treatement of ~$R_b$ ~and ~$R_c$ ~in the fit.

To a very good approximation, the upper bounds result from a combination
of ~$m_t - M_h$ ~correlation given by a fit to all but ~$R_b$ ~
electroweak data and the upper bound on ~$m_t$ ~obtained from ~$R_b$ ~and
the direct measurement of ~$m_t$. ~The SLD ~$A_{LR}^e$ ~result has
important impact on ~
$m_t - M_h$ ~correlation in the direction of lowering the values of ~
$M_h$ ~for given ~$m_t$ ~(as seen in Fig. 3a).

We also note that inclusion of the SLD result in the fits gives lower
value of ~$\alpha_s(M_Z)$. ~This follows from the fact that the  ~$m_t
- M_h$ ~values required by the value of ~$A_{LR}^e$ ~tend to increase ~
$\Gamma_{h}$, ~i.e. a lower value of ~$\alpha_s(M_Z)$ ~is now needed.

Finally one can add a few remarks related to the new values
of ~$R_b$ ~and ~$R_c$. ~Following earlier references \cite{BV} we
have discussed in the original version of the
paper the possibility of new physics in ~$R_b$ ~and studied the
upper bounds on ~$M_h$ ~in a toy model which reproduces all the results
of the SM  and has an {\it ad hoc} correction to the ~
$Z\rightarrow\overline bb$ ~width, so that ~$R_b\approx0.22$. Then the fitted
value of ~$\alpha_s(M_Z)=0.108$ ~and the observable ~$R_b$ ~becomes
irrelevant (its contribution to ~$\chi^2\approx0$) ~from the point of view
of the limits on ~$M_h$. ~In this toy model the limits on ~$M_h$ ~were
similar to the SM limits without ~$R_b$ ~in the fit (somewhat weaker due to
the smaller value of ~$\alpha_s(M_Z)$). ~With the new data the discussion
remains valid provided we use the value of ~$R_b$ ~extracted under the
assumption of ~$R_c$ ~fixed to the SM value and disregard the new ~$R_c$ ~
measurement. If, however, we include both measurements, ~$R_b$ ~and ~
$R_c$, ~and consider a toy model with new physics in both variables, so that ~
$R_c\approx0.1540$ ~and ~$R_b\approx0.222$ ~then the fitted value of ~
$\alpha_s(M_Z) = 0.178\pm0.005$. ~This is easy to understand: with ~
$\Gamma_{Z^0\rightarrow hadr}=1744.8\pm3.0$ ~the sum ~
$\Gamma_{Z^0\rightarrow\overline bb} + \Gamma_{Z^0\rightarrow\overline cc}$ ~
is now much smaller than in the SM and we get ~
$\Gamma_{Z^0\rightarrow hadr} -
\Gamma_{Z^0\rightarrow\overline bb} - \Gamma_{Z^0\rightarrow\overline cc}$ ~
large. Of course, also in this toy model the observables ~$R_b$ ~and ~
$R_c$ ~become irrelevant for the bounds on ~$M_h$ ~and we get them similar
to the SM fit without ~$R_b$ ~in the fit. However, the large value of the
fitted ~$\alpha_s(M_Z)$ ~casts doubts on the measurement of ~$R_c$, ~
unless new physics also alters the light quark sector.
(We thank H.E. Haber and C. Wagner for the discussion on this point.)

Our results, in part which overlaps, are in very good agreement with
recent fits presented in \cite{ELLIS}.

\vskip 0.3cm P.Ch. would like to thank Max Planck Institute in Munich
for hospitality.

\newpage

\noindent {\bf FIGURE CAPTIONS}
\vskip 0.5cm

\noindent {\bf Figure 1.}

\noindent Contours of ~$\Delta\chi^2=1$ ~and ~$\Delta\chi^2=4$ ~
for the ~$\chi^2$  ~fit to all LEP, SLD and Fermilab data.
\vskip 0.3cm

\noindent {\bf Figure 2.}

\noindent Contours of ~$\Delta\chi^2=1$ ~and ~$\Delta\chi^2=4$ ~
for the ~$\chi^2$  ~fits without the asymmerties measured at SLD ~
($A_{LR}^e$, ~$A_{LR}^b$, ~$A_{LR}^c$): \\
a) without ~$R_b$, ~$R_c$,\\
b) with ~$R_b=0.2206\pm0.0016$ ~($R_c=0172$ ~fixed) included.\\
In both figures results of
the fits without/with the Fermilab value for ~$m_t$ ~
included are marked as ~$1,2\sigma_{-t}$/$1,2\sigma_{+t}$ ~
In  the right bottom corner of Fig. 2a the line of
constant ~$A_{LR}^e=0.1551$ ~as measured at SLAC is also shown.
\vskip 0.3cm

\noindent {\bf Figure 3.}

\noindent As in Fig. 2 but with the SLD result for ~$A_{LR}^e$ ~included.
\vskip 0.3cm

\end{document}